\documentclass[twocolumn,preprintnumbers,amsmath,amssymb]{revtex4}

\usepackage{graphicx}
\usepackage{dcolumn}
\usepackage{bm}
\usepackage{color}
\usepackage{setspace}

\begin{document}


\title{Universal scaling for the spin-electricity conversion on surface states of topological insulators}

\author{K. T. Yamamoto$^{\, 1}$}
\thanks{K. T. Yamamoto and Y. Shiomi contributed equally to this work.}
\author{Y. Shiomi$^{\, 1,2}$}
\thanks{Corresponding author}
\email{shiomi@imr.tohoku.ac.jp}
\author{Kouji Segawa$^{\, 3,4}$}
\author{Yoichi Ando$^{\, 3,5}$}
\author{E. Saitoh$^{\, 1,2,6,7}$}

\affiliation{$^{1}$
Institute for Materials Research, Tohoku University, Sendai 980-8577, Japan }
\affiliation{$^{2}$
Spin Quantum Rectification Project, ERATO, Japan Science and Technology Agency, Aoba-ku, Sendai 980-8577, Japan}
\affiliation{$^{3}$
Institute of Scientific and Industrial Research, Osaka University, Ibaraki, Osaka 567-0047, Japan
}
\affiliation{$^{4}$
Department of Physics, Kyoto Sangyo University, Kyoto 603-8555, Japan.
}
\affiliation{$^{5}$
Institute of Physics II, University of Cologne, D-50937 Cologne, Germany
}
\affiliation{$^{6}$
WPI Advanced Institute for Materials Research, Tohoku University, Sendai 980-8577, Japan
}
\affiliation{$^{7}$
Advanced Science Research Center, Japan Atomic Energy Agency, Tokai 319-1195, Japan
}

\date{\today}

\begin{abstract}
We have investigated spin-electricity conversion on surface states of bulk-insulating topological insulator (TI) materials using a spin pumping technique. The sample structure is Ni-Fe$\mid$Cu$\mid$TI trilayers, in which magnetic proximity effects on the TI surfaces are negligibly small owing to the inserted Cu layer. Voltage signals produced by the spin-electricity conversion are clearly observed, and enhanced with decreasing temperature in line with the dominated surface transport at lower temperatures. The efficiency of the spin-electricity conversion is greater for TI samples with higher resistivity of bulk states and longer mean free path of surface states, consistent with the surface spin-electricity conversion.  
\end{abstract}

\maketitle

Injection and detection of non-equilibrium spins are key techniques in the field of spintronics \cite{spin-current}. A powerful method to inject spins is spin pumping. Spin pumping enables dynamical spin injection from a ferromagnet into an adjacent non-magnetic metal, which is induced by coherent precession of magnetization at ferromagnetic resonance (FMR) \cite{saitoh}. In spin pumping experiments, bilayers comprising ferromagnetic Ni$_{81}$Fe$_{19}$ (permalloy, Py) and nonmagnetic Pt have been studied as a typical system  \cite{saitoh, azevedo, costache, ando, mosendz}. Though Pt has been widely used for spin detection owing to its strong spin-orbit interaction, search of more efficient spin detectors is one of the urgent issues in the spintronics field \cite{niimi, matsuno}.    
\par

A topological insulator (TI) is a promising material for spintronics application because of its potential for highly efficient spin-electricity conversion \cite{ralph, yabin}. Topological insulators are a state of quantum matter \cite{TI-review1, TI-review2, TI-review3}, in which the surface is metallic, while the interior is insulating. Spin-electricity conversion on TI materials has recently been investigated using spin pumping for bulk-metallic samples (Bi$_{2}$Se$_{3}$) \cite{Baker, Deorani, Jameli} and also for bulk-insulating ones \cite{shiomi}. In a previous report \cite{shiomi}, some authors of the present paper demonstrated the spin-electricity conversion induced by spin pumping into surface states of TIs, Bi$_{1.5}$Sb$_{0.5}$Te$_{1.7}$Se$_{1.3}$ (BSTS) \cite{taskin, ren} and Sn-doped Bi$_{2}$Te$_{2}$Se (SnBTS) \cite{ren2} in contact with Py. Since millimeter-thick TI samples were used in \cite{shiomi}, the inverse spin Hall signal from bulk carriers are neglected and the observed spin-electricity conversion signal was safely ascribed to a surface contribution. The sign of the generated electric signals is consistent with the spin-electricity conversion on the topological surface state, whereas the opposite sign is expected for co-existing Rashba surface states \cite{SHong}. On the surface states of TIs, since spin direction and electron-flow direction have one-to-one correspondence (the spin-momentum locking), injected spins are converted into electric currents along an in-plane direction on the surface, when the bulk state is sufficiently insulating. Although highly efficient spin-electricity conversion has been reported for bulk-metallic TIs \cite{ralph, yabin, Jameli}, the reported value for bulk-insulating BSTS is $\sim 0.01$\% \cite{shiomi, shiraishi}.
\par

Though the spin-electricity conversion on the topological surface states was demonstrated using a spin pumping technique \cite{shiomi}, it remains unclear how the produced electric signal is related to surface transport properties of TI materials. In this manuscript, we study spin-electricity conversion induced by spin pumping for several SnBTS samples whose bulk insulating properties are ideal for its detailed study. By inserting a thin Cu layer between Py and TI layers, magnetic proximity effects to TI surfaces are negligible. The efficiency for the observed spin-electricity conversion are found to be greater for TI materials with higher bulk resistivity and longer surface mean path. This result shows that the spin-electricity conversion takes places at the surface state and also that its efficiency strongly depends on surface transport properties.  
\par

\begin{figure}[t]
\begin{center}
\includegraphics[width=8.5cm]{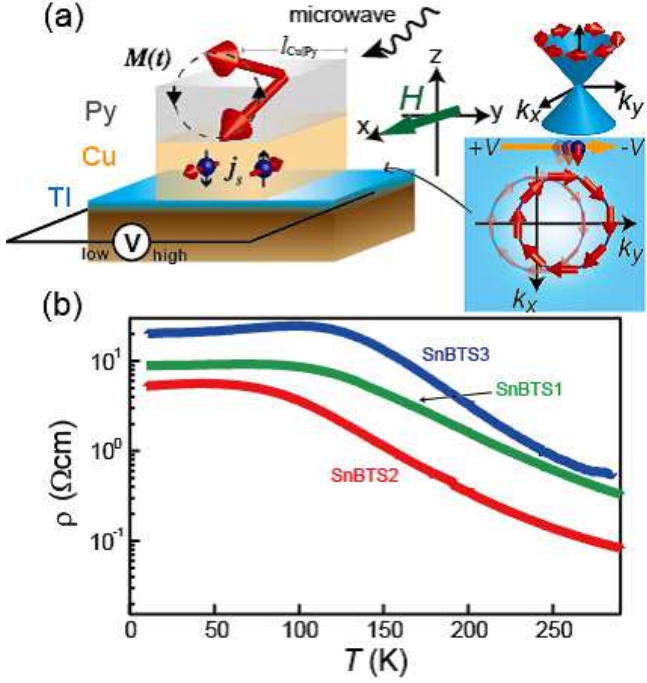}
\caption{(Color online.) (a) Experimental setup of spin-electricity conversion induced by spin pumping into topological insulators (TIs). ``high" and ``low" indicate input codes of the nanovoltmeter (Keithley 2182A). On the topological surface state, spin polarization produces spin-electricity conversion voltage in the Hall direction. The origin of this electric signal is a shift of the surface Fermi surface induced by spin injection. (b) Temperature ($T$) dependence of the in-plane resistivity ($\rho$) for three Sn-doped Bi$_{2}$Te$_{2}$Se samples (SnBTS1, SnBTS2, and SnBTS3). } 
\label{fig1}
\end{center}
\end{figure}

We used SnBTS to investigate the spin-electricity conversion effect on topological surface states. The single crystals of $0.5$\% Sn-doped Bi$_{2}$Te$_{2}$Se were synthesized by a Bridgman method \cite{ren2}. Since the bulk state of SnBTS is more insulating than that of BSTS \cite{ren2, cava}, SnBTS has been considered as an ideal system for transport study of Dirac surface states \cite{cava}. Three SnBTS samples (SnBTS1 whose size is $3.3\times1.2\times0.4$ mm$^{3}$, SnBTS2 $3.3\times2.2\times0.3$ mm$^{3}$, and SnBTS3 $3.0\times1.5\times0.7$ mm$^{3}$) were used in this study. These samples were cut from one boule of SnBTS crystals.
\par

The experimental setup of the spin-pumping measurement is almost the same as that in our previous report \cite{shiomi}. Schematic illustrations of the sample and experimental setup are shown in Fig. \ref{fig1}(a). A $5$-nm-thick Cu film and $25$-nm-thick Py film were evaporated on the middle part of a cleaved surface of TI samples in a high vacuum. The roughness of TI surfaces measured by atomic force microscopy is about $1$ nm. The length of Cu and Py films ($l_{\rm Cu \mid Py}$) is $0.5$ mm. In the spin-pumping measurement, magnetization dynamics in Py was excited by a microwave magnetic field on a coplanar-type waveguide in an inplane static magnetic field ($H$). We used a commercial network analyzer as a microwave source. Microwave frequency was kept at $5$ GHz and the power of incident microwave was amplified $1000$ times by a commercial microwave amplifier. While sweeping the external magnetic field, FMR spectrum in Py and electromotive force arising between the ends of TI samples were recorded simultaneously using the network analyzer and a nanovoltmeter, respectively. The measurements were conducted at low temperatures down to $10$ K in a probe station. 
\par

\begin{figure}[t]
\begin{center}
\includegraphics[width=8.5cm]{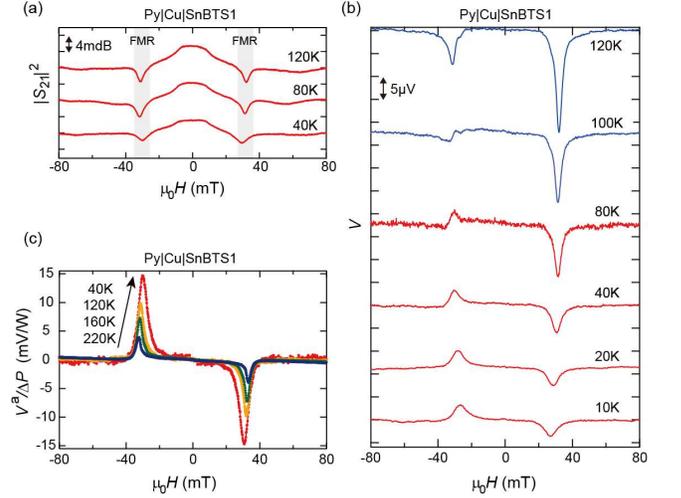}
\caption{(Color online.) (a) Magnetic field ($H$) dependence of microwave transmittance, $|S_{21}|^{2}$, for Py$\mid$Cu$\mid$SnBTS1 at several temperatures. The data is shifted vertically just for clarity. (b) Magnetic field ($H$) dependence of voltage signal ($V$) arising around FMR magnetic-fields of Py for Py$\mid$Cu$\mid$SnBTS1 at various temperatures. The data is shifted vertically just for clarity. (c) $H$ dependence of the antisymmetric part of $V$ ($V^{a}$) normalized by the resonance absorption power ($\Delta P$) for Py$\mid$Cu$\mid$SnBTS1 at some temperatures. } 
\label{fig2}
\end{center}
\end{figure}

Figure \ref{fig1}(b) shows temperature ($T$) dependence of resistivity, $\rho$, for the TI samples. $\rho$ at room temperature is $0.05$ ${\rm \Omega cm}$ - $1$ ${\rm \Omega cm}$ and increases with decreasing $T$, which indicates that the bulk carriers are compensated in all the TI samples. Using the activation law above $200$ K, values of energy gap are estimated to be $80$ meV - $100$ meV, similar to a reported value ($65$ meV) \cite{ren2}. At low temperatures below $\sim$$100$ K, the resistivity begins to decrease, which is unusual for traditional semiconductors. Metallic surface conduction is dominant in the low-$T$ region.    
\par

We performed spin pumping experiments for the TI samples attached with Cu and Py films. Figure 2(a) shows the magnetic field ($H$) dependence of microwave transmittance, $|S_{21}|^{2}$, for Py$\mid$Cu$\mid$SnBTS1 at several temperatures. Clear dips which correspond to ferromagnetic resonance (FMR) in Py were observed at each temperature around $\pm 30$ mT. The magnitude of resonance absorption slightly decreases with decreasing $T$ because of an increase in the microwave loss in our microwave circuit. \par

\begin{figure}[t]
\begin{center}
\includegraphics[width=8.5cm]{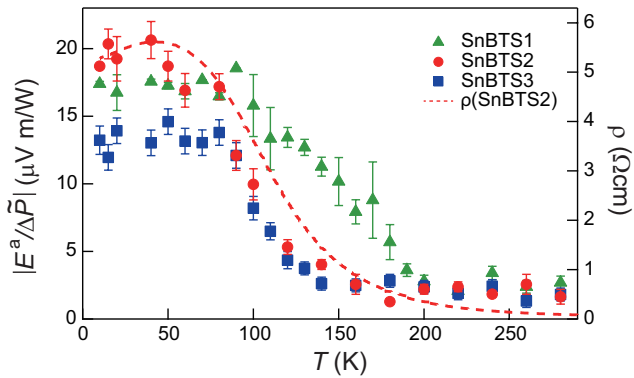}
\caption{(Color online.) Temperature ($T$) dependence of the antisymmetric part of the electric field, $E^{a} (=V^{a}/l_{\rm Cu \mid Py})$, divided by the resonance absorption power per unit area ($\Delta \tilde{P}$) for the Py$\mid$Cu$\mid$SnBTS samples. The absolute values of $E^{a}/\Delta \tilde{P}$ are plotted. The $T$ dependence of resistivity ($\rho$) for SnBTS2 is shown for comparison (dotted curve).  }
\label{fig3}
\end{center}
\end{figure}

Figure \ref{fig2}(b) shows the magnetic field ($H$) dependence of voltage signals arising at FMR magnetic fields for Py$\mid$Cu$\mid$SnBTS1 at various temperatures. At $120$ K, the voltage peaks are clearly observed at FMR magnetic fields ($\pm H_{FMR}$) of Py. The sign of the voltage peaks is negative both at $+H_{FMR}$ and $-H_{FMR}$, and their magnitudes are different. At high temperatures, the Seebeck effect of bulk carriers independent of magnetic fields dominates the peak signals \cite{shiomi}. An origin of this Seebeck voltage is an inplane temperature gradient due to the inevitable heating effects at FMR \cite{shiomi}. As temperature decreases, the magnitude of the Seebeck voltage decreases since the excitation of bulk carriers is suppressed at lower temperatures. Below $80$ K, the sign reversal between $+H_{FMR}$ and $-H_{FMR}$ is observed as shown in Fig. \ref{fig2}(b), indicating that the spin-electricity conversion signal is dominant. The peak sign at $+H_{FMR}$ is negative, consistent with that reported in \cite{shiomi}. 
\par

The magnitude of the spin-electricity conversion voltage increases with decreasing temperature, consistent with the dominant surface conduction at lower temperatures. Figure \ref{fig2}(c) shows the $H$ dependence of the antisymmetric part of the voltage peak $|V^{a}| \equiv |V(H)-V(-H)|/2$ divided by resonance absorption power $\Delta P$ for Py$\mid$Cu$\mid$SnBTS1. Here $\Delta P$ is calculated from the microwave transmittance data for the same sample in Fig. \ref{fig2}(a) \cite{shiomi}. As shown in Fig. \ref{fig2}(c), the magnitude of $|V^{a}/\Delta P|$ monotonically increases with decreasing $T$ from $220$ K to $40$ K for Py$\mid$Cu$\mid$SnBTS1.  
\par

\begin{figure}[t]
\begin{center}
\includegraphics[width=8.5cm]{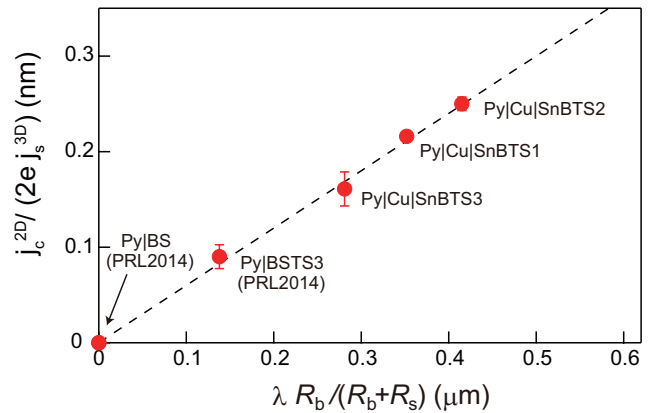}
\caption{(Color online.) The inverse Edelstein effect length, $j_{c}^{2D}/(2ej_{s}^{3D})$, as a function of $\lambda R_{s}/(R_{s}+R_{b})$ for various samples. $j_{c}^{2D}$ and $j_{s}^{3D}$ indicate the electric current density generated on the TI surface and the spin current density injected from Py, respectively. Here, $e$, $\lambda$, $R_{s}$, and $R_{b}$ denote the elementary charge, the mean-free path on the surface, and the sheet resistances for the surface and bulk parts, respectively. The voltage-peak magnitude at the lowest temperature was used to calculate $j_{c}^{2D}$. The experimental results for Py$\mid$BS and Py$\mid$BSTS3 reported in [\onlinecite{shiomi}] are also plotted in this figure. }
\label{fig4}
\end{center}
\end{figure}

Figure \ref{fig3} summarizes the $T$ dependence of the antisymmetric signals for all the samples. Here, the antisymmetric signal $|V^{a}/\Delta P|$ is normalized by the sample size; the produced electric field $E^{a} \equiv V^{a}/l_{\rm Cu \mid Py}$ divided by the resonance absorption power per unit area $\Delta {\tilde P}$ is plotted in Fig. \ref{fig3}. As shown in Fig. \ref{fig3}, $|E^{a}/\Delta \tilde{P}|$ for the three samples is small and almost constant above $200$ K. The magnitudes of $|E^{a}/\Delta \tilde{P}|$ are similar for all the samples in the high $T$ range. This small constant signal at high temperatures may result from ferromagnetic transports in the Py layer \cite{shiomi}. With decreasing temperature below $150$ K - $200$ K, $|E^{a}/\Delta \tilde{P}|$ begins to increase rapidly for the Py$\mid$Cu$\mid$SnBTS samples. This sharp enhancement of $|E^{a}/\Delta \tilde{P}|$ at low temperatures is clearly correlated with the resistivity increase at low temperatures, as shown by the dotted curve in Fig. \ref{fig3}. As the bulk resistivity increases at lower temperatures, the spin polarization accumulated near the TI surface increases and greater spin-electricity conversion signals may appear. The magnitude of $|E^{a}/\Delta \tilde{P}|$ at the lowest temperature is similar for all the samples, whereas the resistivity values are rather different among them [Fig. \ref{fig1}(b)].    
\par

In Fig. \ref{fig4}, we analyze the efficiency of the spin-electricity conversion in terms of the mean-free path on the surface $\lambda$ for each sample. The inverse Edelstein effect length \cite{sanchez1, sanchez2, IEEtheory} for the Py$\mid$Cu$\mid$SnBTS samples and also for Py$\mid$Bi$_{2}$Se$_{3}$ (Py$\mid$BS) and Py$\mid$Bi$_{1.5}$Sb$_{0.5}$Te$_{1.7}$Se$_{1.3}$3 (Py$\mid$BSTS3) reported in \cite{shiomi} (see also \cite{note}) is calculated by dividing the surface electric current density $j^{2D}_{c} \equiv \{ E^{a}({\rm 10K}) - E^{a}({\rm 293K})\}/R_{t}$ by the spin current density $j^{3D}_{s}$ \cite{ando}, where the electric current originating from the ferromagnetic transports in the Py layer ($E^{a}({\rm 293K})/R_{t}$) is subtracted. The obtained $j_{c}^{2D}/(2ej_{s}^{3D})$ is plotted against $\lambda R_{b}/(R_{b}+R_{s})$ for each sample in Fig. \ref{fig4}. Here, $R_{t}$, $R_{s}$, and $R_{b}$ are the total sheet resistance for the Py$\mid$Cu layer and the TI surface in parallel, the surface sheet resistance, and the bulk sheet resistance, respectively. $\lambda$ is estimated from the reported Fermi wavenumber $k_{F}$ ($5.9 \times 10^{-2}$ \AA$^{-1}$ for Sn-BTS \cite{ren2, cava} and $1.0 \times 10^{-1}$ \AA$^{-1}$ for BSTS \cite{arakane}) and $R_{s}$ estimated for each sample \cite{culcer}, and $R_{b}$ and $R_{s}$ are separated from each other by fits to the $\rho$-$T$ curve using the three-dimensional variable range hopping formula ($\sim T^{-1/4}$) in a low-$T$ range \cite{ren}. For the bulk-metallic TI samples (BS), $R_{b}/(R_{b}+R_{s}) \approx 0$, and the spin-electricity conversion signal is not observed \cite{shiomi} (Fig. \ref{fig4}). For the bulk-insulating TI samples, by contrast, both $j^{2D}_{c}$ and $R_{b}/(R_{b}+R_{s})$ exhibit sizable values. As shown in Fig. \ref{fig4}, a clear linear relation is observed, which shows that the spin-electricity conversion efficiency is greater for TI samples with higher bulk resistivity and longer surface mean-free path. Experimental results for Py$\mid$BSTS1 and Py$\mid$BSTS2 in literature \cite{shiomi} are not shown in Fig. \ref{fig4}, since the antisymmetric parts of the voltage peaks are not discerned above $100$ K due to very large Seebeck voltages \cite{shiomi}.    
\par

The linear dependence of the inverse Edelstein effect length on mean free path on the surface is consistent with the theoretical prediction \cite{sanchez1, sanchez2, IEEtheory}. The relation between the generated electric current and the spin polarization $\langle \sigma   \rangle$ is given by $j_{c}^{2D} \approx 2e v_{F} \langle \sigma   \rangle$ \cite{burkov}, where $v_{F}$ is the Fermi velocity for the surface state. The spin polarization on the surface is produced by injection of the spin current $j_{s}^{3D}$ from Py, following $\langle \sigma \rangle / \tau = j_{s}^{3D} R_{b}/(R_{b}+R_{s})  $ ($\tau$ being the scattering relaxation time for the surface) \cite{shiomi}. Hence, we obtain $j_{c}^{2D}/j_{s}^{3D} \approx 2e \lambda R_{b}/(R_{b}+R_{s}) $. This relation well explains the experimental results in Fig. \ref{fig4}. It is interestingly noted that the slope of the linear fit in Fig. \ref{fig4} corresponds to the spin injection efficiency $\eta$ \cite{shiomi}. The obtained value is $\eta \approx 6.0 \times 10^{-4}$ ($0.06$\%), which is comparable to the values reported for bulk-insulating BSTS samples \cite{shiomi, shiraishi}.   
\par

In summary, we measured spin-electricity conversion induced by spin pumping into Sn-doped Bi$_{2}$Te$_{2}$Se at low temperatures. To prevent magnetic proximity effects from ferromagnetic layers, a thin Cu layer was inserted between Py and TI. At FMR in the Py, spin-electricity conversion voltage was observed and enhanced with decreasing temperature. The spin-electricity conversion efficiency at low temperatures is found to increase with increasing magnitudes of bulk resistivity and surface mean-free path, following a single scaling law. This result is consistent with theories on spin-electricity conversion on the topological surface state.
\par

K.T.Y. acknowledges the support through the Leading Graduates Schools Program (Tohoku University ``MD program")  by the MEXT. Y.S. would like to acknowledge the support from the Motizuki Fund of Yukawa Memorial Foundation. This work was supported by JSPS (KAKENHI No. 25220708 and the Core-to-Core program ``International research center for new-concept spintronics devices") and MEXT (Innovative Area gNano Spin Conversion Scienceh (No. 26103005)).
\par

\end{document}